\documentclass[12pt]{article}

\usepackage{graphicx}
\usepackage{amssymb,latexsym}
\usepackage{amsmath}
\usepackage[dvips]{epsfig}
\begin{document}

\title{\bf{Surface effects on the statistics of the local density of states in metallic
nanoparticles: manifestation on the NMR spectra.}}
\author{Jos\'{e} A. Gasc\'{o}n$^1$  \\
Department of Chemistry, Yale University\\
P.O. Box  208107, New Haven, CT 06520-8107, USA\\
\\
Horacio M. Pastawski\\
Facultad de Matem\'{a}tica Astronom\'{\i}a y F\'{\i}sica, Universidad\\
Nacional de C\'{o}rdoba, Ciudad Universitaria, 5000 C\'{o}rdoba, Argentina}

\maketitle

\footnotetext[1]{e-mail: jose.gascon@yale.edu (author to whom correspondence should
be addressed)}

\begin{abstract}

\setlength{\baselineskip}{0.3in}

In metallic nanoparticles, shifts in the ionization energy of surface atoms with 
respect to bulk atoms can lead to surface bands. Within a simple Tight Binding model 
we find that the projection of the electronic density of states on these sites 
presents two overlapping structures. One of them is characterized by the level 
spacing coming from bulk states and the other arises from the surface states. In very 
small particles, this effect contributes to an over-broadening of the NMR absorption 
spectra, determined by the Knight shift distribution of magnetic nuclei. We 
compare our calculated Knight shifts with experiments on aluminum nanoparticles, 
and show that the deviation of the scaling law as a function of temperature and 
particle size can be explained in terms of surface states.
  
\end{abstract}

\setlength{\baselineskip}{0.3in}

\section{Introduction}

Several NMR experiments on metallic nanoparticles have shown \cite{Halp-Brom,Slichter,V.D.Klink} 
quantum size effects: the absorption spectra becomes quite broad and asymmetric when either 
temperature or particle size diminishes. Studies of relaxation time \cite{V.D.Klink} 
$T_{1\text{ \ }}$ indicate that the spectra is inhomogeneous, implying phenomena of a local
nature. This inhomogeneity has been interpreted \cite{Efetov} considering a strictly local 
description for the electron polarization in the presence of a magnetic field $H$. Given the 
Bohr magneton, $\mu _{\text{B}}$, and $N(\varepsilon _{{\rm F}},${\bf r}$)$, the local density 
of {\it s} states at the Fermi level $\varepsilon _{{\rm F}}$, the {\it local} Pauli
susceptibility is written as: 
\begin{equation}
\chi _{\text{p}}({\bf r)}=\mu _{\text{B}}^{2}N(\varepsilon _{{\rm F}},{\bf r} 
).  \label{Eq. 1}
\end{equation}
This equation expresses that, due to the finite size effects and
interference phenomena proper of the mesoscopic systems, the electronic spin
polarization is {\it inhomogeneous}. The hyperfine coupling between nuclear
spins and electronic spins produces different shifts in the resonance
frequency for {\it each} nucleus in the nanoparticle:
\begin{equation}
\Delta \omega ({\bf r)\propto }\chi _{\text{p}}({\bf r)}H.  \label{Eq. 2}
\end{equation}
Therefore, an inhomogeneously broadened NMR\ absorption line is the key to
indirectly measure the fluctuations of the local density of states (LDOS).

In previous works \cite{Pastaw-Gasc} we have calculated the NMR line shape
in metallic nanoparticles within a Tight Binding model with $M$ orbitals
accounting for the fluctuations of the LDOS. A remarkable outcome,
consistent with many experimental results \cite{Halp-Brom,Volokitin}, is a
universal scaling behavior of the line position and shape with respect to
the variation of the thermal energy $k_{{\rm B}}T$ and the mean spacing level 
$\Delta \cong 1/\left[ M\times N_{0}(\varepsilon _{{\rm F}})\right] $. This
results in a universal scaling law for the Knight shifts and the Pauli susceptibility 
\cite{Volokitin} whose relevant parameter is 
\begin{equation}
 \alpha = k_{{\rm B}}T/\Delta. 
\label{alpha}
\end{equation}
However, in very small crystalline particles, the experimental line shape
presents an anomalous over-broadening at intermediate temperatures \cite{Brom}
. In addition, a.c. conductivity measurements show that tunneling among
particles plays a relevant role in the electronic properties \cite{Brom2}. This suggests
the need to explore for additional sources of fluctuations on the statistics of energy levels, 
particularly the effects of surface states. It is known that self
consistent calculations of metal surfaces \cite{Guevara,Ganduglia} give
different ionization energies for surface and bulk sites. This fact is
expected both in metallic particles with clean surfaces and as well as those
with chemisorbed atoms. In our previous calculation we made no attempt to
consider surface states. However the natural connectivity of the surface
orbitals caused slight departures from the universal scaling law \cite{Pastaw-Gasc}.

In the present work we use a Tight Binding Hamiltonian with an {\it ad hoc} 
shift in the surface site energies \cite{Kalkstein,Ganduglia}. This
produces surface states, and as a consequence, large fluctuations on
the LDOS distribution. According to the Eqs. \ref{Eq. 1} and \ref{Eq. 2},
this leads to a modified distribution of Knight shifts detected as a
wider NMR absorption lines. We compare our calculations of the Knight shift
as a function of the scaling parameter with experimental values on
aluminum nanoparticles and show that the observed deviation of the scaling
law is a manifestation of surface effects.

\section{Model Hamiltonian}

A metal particle with $M$ atoms can be modeled with the Hamiltonian

\begin{equation}
{\cal H}=\sum_{i=1}^{M}E_{i}c_{i}^{+}c_{i}+\sum_{j>i}^{M} 
\sum_{i=1}^{M}(V_{ji}c_{j}^{+}c_{i}+V_{ij}c_{i}^{+}c_{j}),  \label{H}
\end{equation}
where $E_{i}$ is the energy of an $s$ state centered at site $i$ of a cubic
lattice.{\it \ }$V_{ij}\equiv V$ is the kinetic energy involved in hopping
between nearest neighbors sites $i$ and $j$. To represent shape and crystal
inhomogeneities, the sites energies are taken in the range $-W/2$ and $W/2$
(Anderson's disorder)$.$ We define \emph{surfaces sites} as the sites which
being at the surface of the cube, have an additional energy shift $U$.
The rest of the sites are \emph{bulk sites}, even though they may belong to the
faces of the cube with un-shifted site energies. This allows to model different surface to bulk
ratios maintaining $M$ constant. Tunneling among particles produces an
inhomogeneous broadening \cite{GLBE} of the atomic energy levels ($ 
E_{i}\rightarrow E_{i}-$i$\Gamma _{i}$), taking its higher values $\Gamma
_{s}$ at particular sites (\emph{contacts}) at the surface. Assuming that $ 
\Gamma _{i}\ll V,$ and neglecting localization effects we can assume $\Gamma
_{i}\equiv \eta _{0}\approx \Gamma _{s}/M$ for all sites in the
nanoparticle. As shown in the Ref. 4 finite temperature effects are
included by an additional broadening on the energy levels, $\eta =\eta
_{0}+k_{{\rm B}}T$.  Properties of the single particle excitation spectrum are contained in the retarded (advanced) Green's function
\begin{equation}
G_{i,j}^{R(A)}(\varepsilon _{}^{})=\sum_k\frac{a_k(\mathbf{r_i})a_k^{*}(\mathbf{r_j})}{\varepsilon + [(-){\rm i}\eta -E_k]},  \label{GR}
\end{equation}
where $\psi _k({\bf r})=\sum_ia_k(\mathbf{r_i})\varphi _i({\bf r})$ and $E_k$ are the
exact eigenfunctions (molecular orbitals) and eigenenergies for the isolated
particle, respectively. $\eta $ is a natural broadening of the electronic states and the $%
(-)$ sign corresponds to the retarded Green's function. The local density
of states per atom (LDOS) at the $i$-th site is evaluated as:
\begin{equation}
N(\varepsilon,\mathbf{r_i} )=-(2\pi {\rm i})^{-1}[G_{i,i}^R(\varepsilon
)-G_{i,i}^A(\varepsilon )],  \label{Ni}
\end{equation}
from which the relevant contribution to the density of states per unit
volume at the $i$-th nucleus, $N(\varepsilon ,{\bf r}_i),\,$is obtained. 
The evaluation of the local Green's function in Eq.~\ref{GR} is obtained via the 
Matrix Continued Fraction method \cite{Pastaw}. Its basic idea is to exploit the short range
interactions in the Hamiltonian (\ref{H}) by indexing states in a way that
subspaces representing layers interact through nearest neighbor subspaces.
In matrix form:

\begin{equation}
{\cal H=}\left[
\begin{array}{ccccc}
\ddots & \ddots & {\bf 0} & {\bf 0} & {\bf 0} \\
\ddots & {\bf E}_{n-1,n-1} & {\bf V}_{n-1,n} & {\bf 0} & {\bf 0} \\
{\bf 0} & {\bf V}_{n,n-1} & {\bf E}_{n,n} & {\bf V}_{n,n+1} & {\bf 0} \\
{\bf 0} & {\bf 0} & {\bf V}_{n+1,n} & {\bf E}_{n+1,n+1} & \ddots \\
{\bf 0} & {\bf 0} & {\bf 0} & \ddots & \ddots
\end{array}
\right] ,  \label{HM}
\end{equation}

\noindent where ${\bf 0}$'s are null matrices, ${\bf E}$'s in the diagonal represent
intra-layer interactions while the only non-zero off-diagonal matrices ${\bf V}_{n,n\pm 1}$ 
connect nearest neighbor layers. Detailed structure of
the sub-matrices depends on the lattice, for the cubic structure ${\bf V}%
_{n,n\pm 1}=V{\bf 1}$, with ${\bf 1}$ the identity matrix. The local
retarded Green's functions connecting sites $i$ and $j$ within the $n$-th
layers are arranged in a matrix
\begin{equation}
{\bf G}_{n,n}^R\left( \varepsilon \right) =\left[ \left( \varepsilon +{\rm i}%
\eta \right) {\bf 1}{\Bbb -}{\bf E}_{n,n}-{\bf \Sigma }_n^{R\,+}\left(
\varepsilon \right) -{\bf \Sigma }_n^{R\,-}\left( \varepsilon \right)
\right] ^{-1}
\end{equation}
where the matrix self energies ${\bf \Sigma }_n^{R\,+}$ and ${\bf \Sigma }%
_n^{R\,-}$ are calculated in terms of Matrix Continued Fractions (MCF)
defined through the recurrence relations:
\begin{equation}
{\bf \Sigma }_n^{R\,\,\pm }={\bf V}_{n,n\pm 1}\frac{{\bf 1}}{\left(
\varepsilon +{\rm i}\eta \right) {\bf 1}{\Bbb -}{\bf E}_{n\pm 1,n\pm 1}-{\bf %
\Sigma }_{n\pm 1}^{R\,\,\pm }}{\bf V}_{n\pm 1,n},
\end{equation}
which are calculated with the boundary conditions: ${\bf \Sigma }%
_{L}^{+}\equiv {\bf \Sigma }_1^{-}\equiv {\bf 0}$, where $L$ denotes the number of layers.

\textbf{Model parameters.} Parameters that model an $s$-band are \cite{sband}: 
$V\cong 0.9{\rm eV}$, which is
consistent with a bandwidth of $B=12V\cong 0.8$Ry $=11{\rm eV;}$ a shift $ 
U=2V$, consistent with a shift of $0.12$Ry estimated for metal clusters and
a Fermi energy of $\varepsilon _{{\rm F}}=3V+B/2$, which measured from the
band bottom gives a ratio $\varepsilon _{{\rm F}}/B=0.75$, close to typical
values of metals \cite{sband}. In this work we use $\eta =0.05V$, which assumes that
the main contribution to the level broadening comes from tunneling i.e. $ 
\eta _{0}>k_{{\rm B}}T$ which is consistent with the conductivity
measurements \cite{Brom2}.

The statistical distribution of the LDOS is calculated taking an ensemble of
ten Fermi energies in a range $\delta \varepsilon _{{\rm F}}=0.5V$ around $ 
\varepsilon _{{\rm F}}.$ For each energy ten disorder configurations are
considered. Each LDOS $N(\varepsilon _{{\rm F}},{\bf r})$ is normalized to
the bulk value $N_{0}(\varepsilon _{{\rm F}})$, which is evaluated at the
central site of a particle with $15\times 15\times 15$ orbitals. The
normalized LDOS is defined as $x=N(\varepsilon _{{\rm F}},{\bf r} 
)/N_{0}(\varepsilon _{{\rm F}}),$ and occurs with a probability $I(x)$.
According with Eqs. \ref{Eq. 1} and \ref{Eq. 2}, $I(x)$ is also the absorption at
normalized frequency shifts: $x=(\omega -\omega _{0})/(\omega _{K}-\omega
_{0}).\,$Here $\omega _{K}$ is the bulk metal frequency and $\omega _{0}$
the frequency of metal nuclei in dielectric materials (i.e. salt). 
Due to this correspondence, from now on, $I(x)$ will refer indistinctly to either
the distribution of the normalized LDOS at the Fermi energy or the Knight shift NMR spectrum.
Quantum size effects manifest as shifts of the NMR line maximum:
as the particle size or temperature decreases, the shift goes from $x=1$ (metal
bulk) to $x=0$ (salt) while the line broadens and turns asymmetrical 
\cite{Efetov,Pastaw-Gasc}. It is important to emphasize that the simple model used in this work
only attempts to provide physical insight into the quantum size effects as
manifested in the LDOS distribution, as well as a qualitative description of surface
effects.

\section{Results}

To model small surface to volume ratios proper of a big particle, the
surface sites (with mean site energy $U=2V$) were chosen on a single
face of a $7\times 7\times 7$ cube, i.e. $M_{{\rm surf}}=7\times 7=49$ are
surface sites{\it \ }and $M_{{\rm bulk}}=6\times 7\times 7=294$ are bulk
sites. In this case the ratio of level spacing is $\Delta _{{\rm surf.}}/\Delta _{{\rm bulk}}=4$. 
Results for this particle can be extrapolated
to an actual particle with $40\times 40\times 40$ orbitals having
roughly the same surface to bulk ratio. Figure 1 shows the Knight shift
spectrum obtained from the LDOS occurrence distribution at 
surface sites (thick line). 
\begin{figure}[ht]
\begin{center}
\includegraphics*[scale=1.0]{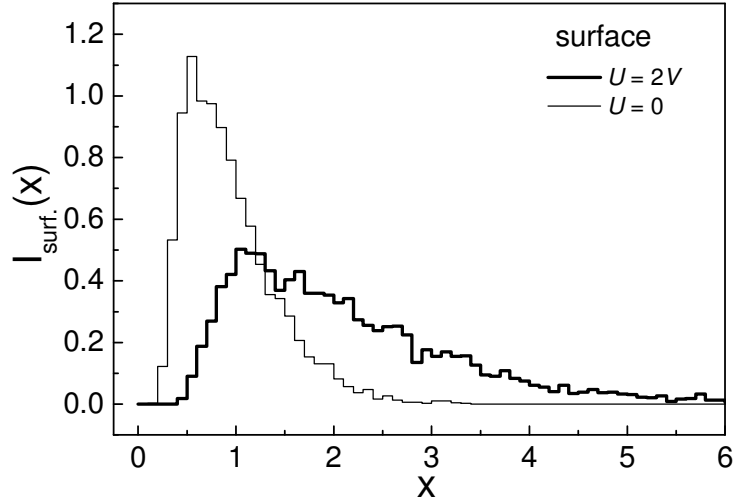}
\caption{Knight shift spectrum of surface sites when $U=2V$ (thick line),
compared with the case with $U=0$ (thin line). $M=7\times 7\times 7,$
$M_{{\rm surf.}}=49$ (corresponding to one side of the cube) and
$M_{{\rm bulk}}=294.$ $W=1V$ and $\eta =0.05V.$}
\end{center}
\end{figure}
It shows that the shift in the site energies produces an over-broadening with
respect to the Knight shift distribution from the same sites for the homogeneous
configuration with $U=0$ (thin line). The over-broadening suggests the
existence of multiple fluctuation scales in the NMR spectrum. To clarify this, 
Figure 2(a) shows the projection of the density of states (DOS) at 
surface sites, $N_{\text{surf.}}(\varepsilon)$. It reveals two structures, one 
corresponding to surface sites with a typical spacing $\Delta _{{\rm surf.}}$ and 
another one with
typical spacing $\Delta _{{\rm bulk}}$ corresponding to bulk sites. Notice
that the structure of bulk states (Fig. 2.b), characterized by a mean level spacing of
$\Delta _{{\rm bulk}}$, enters into the surface DOS as a substructure. These bulk states
have a small weight on the surface band because $U$ provides a
barrier that prevents the mixing between states with the same kinetic energy
($E_{k_{\Vert }}$) parallel to the surface. 

\begin{figure}[ht]
\begin{center}
\includegraphics*[scale=1.0]{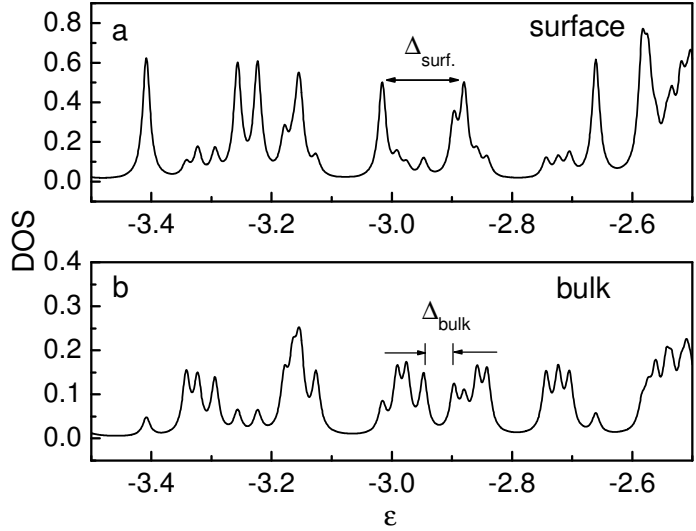}
\caption{DOS projection on a) surface sites{\it \ }with energies $U=2V$ and
b) {\it bulk sites }with energies $U=0.$ $M=7\times 7\times 7,$ $M_{{\rm 
surf.}}=49$ and $M_{{\rm bulk}}=294.$ $W=1V$ and $\eta =0.05V.$}
\end{center}
\end{figure}

The isolated surface and bulk
bands $N_{{\rm surf.}}^{0}(\varepsilon )$ and $N_{{\rm bulk}
}^{0}(\varepsilon ),$ provide a rationale to understand the structure of the
surface DOS, and consequently the multiple fluctuation scales in the NMR spectrum. 
Typically $N_{{\rm surf.}}(\varepsilon )\cong aN_{{\rm surf.} 
}^{0}(\varepsilon )+bN_{{\rm bulk}}^{0}(\varepsilon )$. Within a
perturbative calculation, $a\leq 1-(V^{2}/U)^{2}=3/4$ and $a+b=1$. 
The main peaks in $N_{{\rm surf.}}(\varepsilon )$ (Fig. 2.a) arise from those in 
$N_{{\rm surf.}}^{0}(\varepsilon ).$ They produce a wide range of high LDOS
values with low probability (tail in the $I_{{\rm surf.}}(x)$, Fig. 1). The
valleys of the global structure will produce lower density values with high
probability (peak in the $I_{{\rm surf.}}(x)$). In the same way, the
denser spectrum of the bulk band $N_{{\rm bulk}}^{0}(\varepsilon )$, which
contributes with small weight, provides a lower scale fluctuations to $N_{ 
{\rm surf.}}(\varepsilon )$.
\begin{figure}[ht]
\begin{center}
\includegraphics*[scale=1.0,keepaspectratio=true]{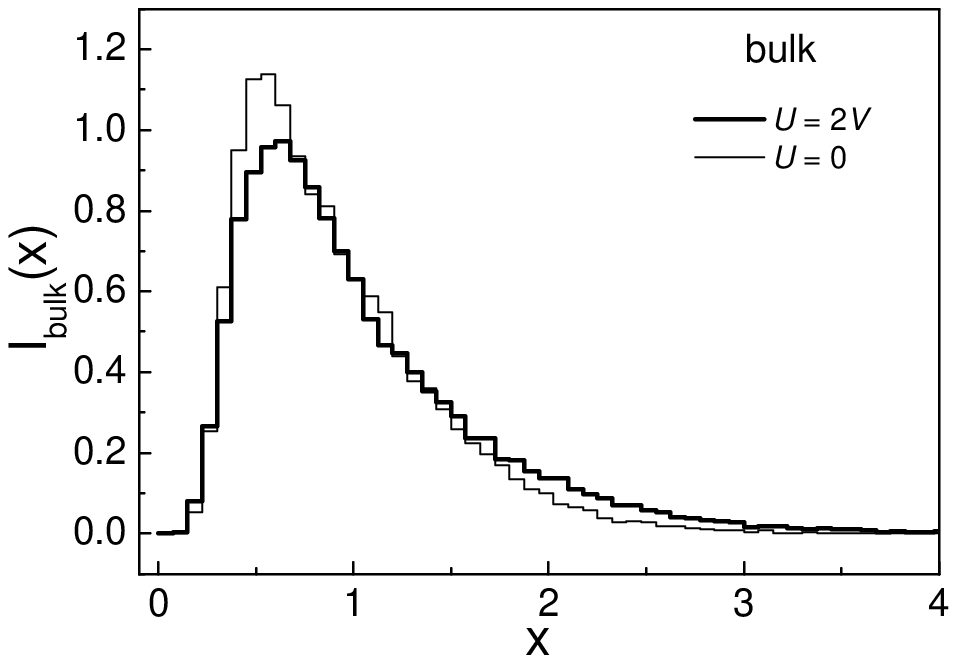}
\caption{Knight shift spectrum of bulk sites when $U=2V$ (thick line),
compared with the case with $U=0$ (thin line). $M=7\times 7\times 7,$ $M_{ 
{\rm surf.}}=49$ and $M_{{\rm bulk}}=294.$ $W=1V$ and $\eta =0.05V.$}
\end{center}
\end{figure}
Again, the peaks and valleys of $N_{{\rm bulk} 
}^{0}(\varepsilon )$ produce an asymmetric distribution $I_{{\rm bulk}}(x)$ 
with a maximum. Therefore, the superposition of these two LDOS will cause the
over-broadening in the distribution of Knight shifts for the surface sites roughly
represented by:
\begin{equation}
I_{{\rm surf.}}(x)\approx \frac{1}{ab}\int I_{{\rm surf.}}^{0}
(\frac{x-x^{\prime }}{a})I_{{\rm bulk}}^{0}(\frac{x^{\prime}}{b})dx^{\prime} 
\end{equation}
Even without attempting a fitting, this shows that the predictions of
numerical results are consistent with the spectral analysis.

\begin{figure}[ht]
\begin{center}
\includegraphics*[scale=1.0,keepaspectratio=true]{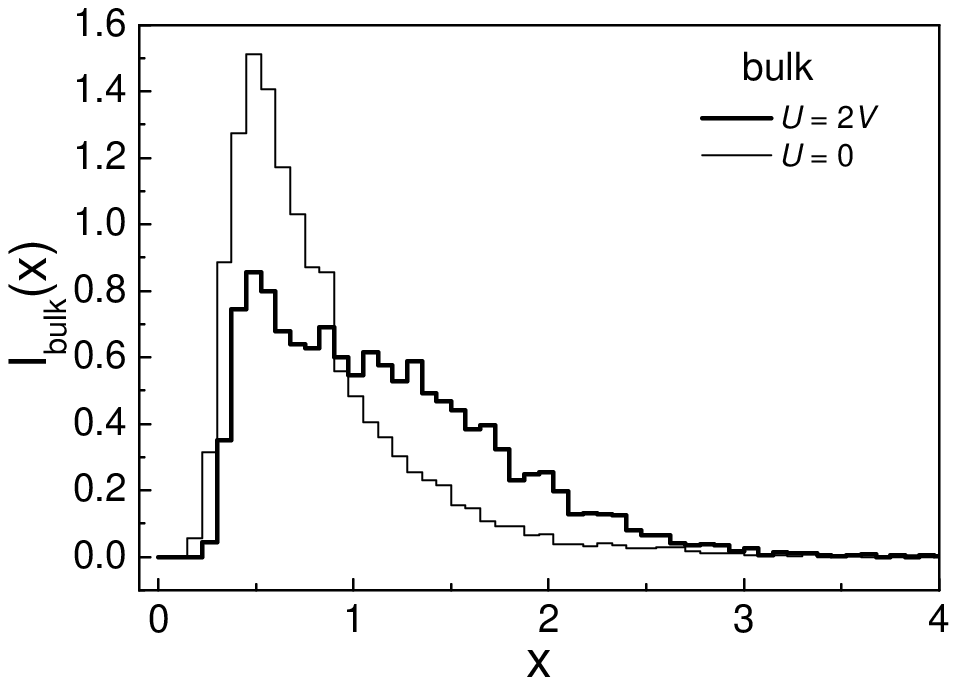}
\caption{Knight shift spectrum of bulk sites setting $U=2V$ (thick line),
compared with the case with $U=0$ (thin line). $M=6\times 6\times 6,$ $M_{ 
{\rm surf.}}=152$ and (corresponding to the six sides of the cube) $M_{{\rm  
bulk}}=64.$ $W=1V$ and $\eta =0.05V.$}
\end{center}
\end{figure}
 
\begin{figure}[ht]
\begin{center}
\includegraphics*[scale=1.0,keepaspectratio=true]{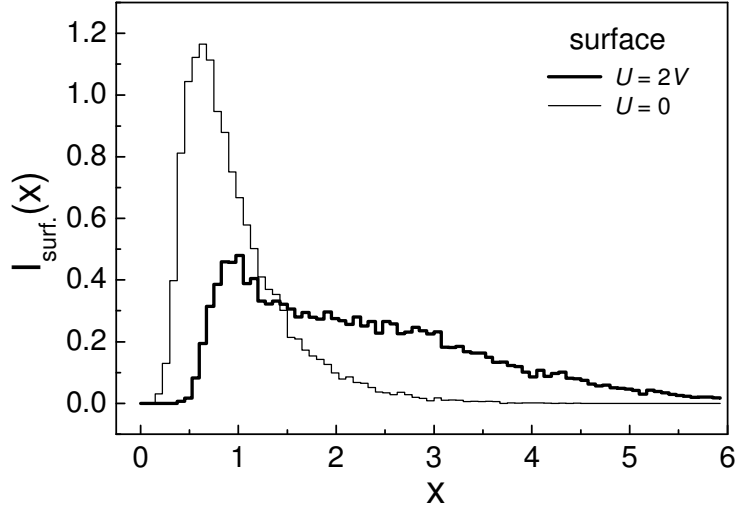}
\caption{Knight shift spectrum of surface sites{\it \ }when $U=2V$ (thick line),
compared with the case with $U=0$ (thin line). $M=6\times 6\times 6,$ $M_{ 
{\rm surf.}}=152$ and $M_{{\rm bulk}}=64.$ $W=1V$ and $\eta =0.05V.$}
\end{center}
\end{figure}
 
\begin{figure}[ht]
\begin{center}
\includegraphics*[scale=1.0]{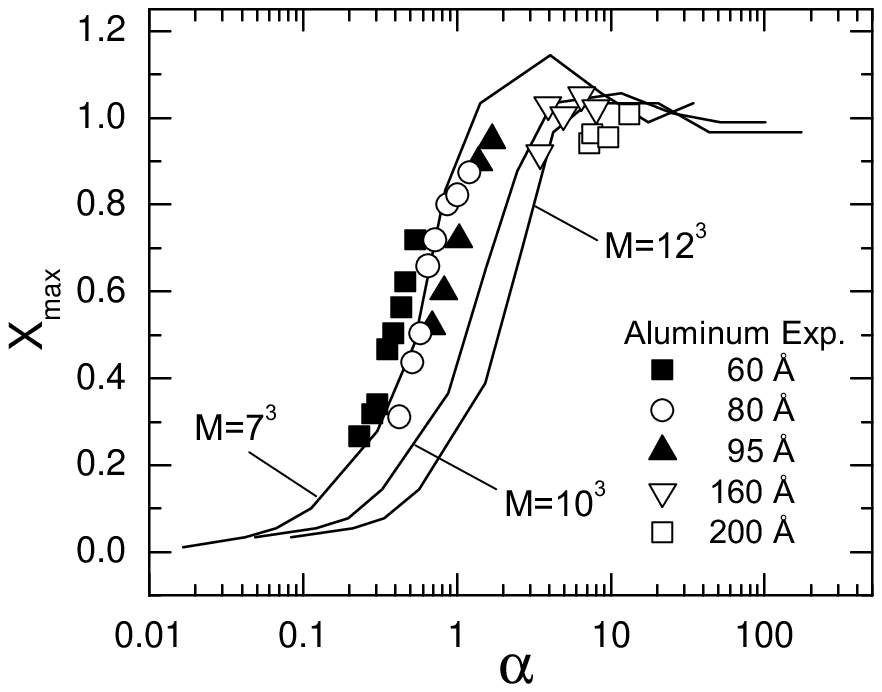}
\caption{Relative Knight shift $x_{\text{max.}}$ \emph{vs.} the scaling parameter $\alpha$.
The solid lines correspond to simulations of various particle sizes with
surface energy $U=2V$ for all faces of the cube.
Experimental values correspond to aluminum nanoparticles \cite{Kobayashi}. In both
simulations and experiments a deviation of the universal scaling law is observed.} 
\end{center}
\end{figure}

In order to evaluate the effect of surface states on bulk NMR signal of big
particles, Figure 3 shows the Knight shift spectrum (thick line) for the
bulk sites together with a case with unperturbed surfaces $(U=0)$
represented by the thin line. $I_{{\rm bulk}}(x)$ does not present much
over-broadening with respect to the homogeneous configuration$.$ The reason
for this can be understood by analyzing the DOS projection on bulk sites.
The relevant scale in $N_{\text{bulk}}(\varepsilon)$ (Fig. 2.b) is $\Delta _{{\rm bulk}}$.
Since surface states enter into the bulk with small weight, and since 
$\Delta _{{\rm surf.}}>\Delta _{{\rm bulk}}$ ($\Delta _{{\rm surf.}}=4\Delta _{{\rm bulk}}$), 
the relevant scale is dictated by spacing of bulk levels. Therefore, the surface band does 
not affect the statistics of the LDOS of bulk sites. 

The previous analysis suggests that if the surface to volume ratio is big,
as in very small particles where $\Delta _{{\rm bulk}}>\Delta _{{\rm surf.} 
}, $ the over-broadening should occur on the LDOS distribution of bulk
sites{\it . }The cluster configuration must be such that the surface sites 
{\it \ }are in higher proportion with respect to the bulk sites{\it . }This
situation can be obtained for a structure with $6\times 6\times 6$ orbitals,
taking all faces of the cube with shifted site energies $U=2V$ ($M_{{\rm surf. 
}}=152$ and $M_{{\rm bulk}}=64$). In this case $\Delta _{{\rm bulk} 
}=3.5\Delta _{{\rm surf.}}$. Figure 4 shows the Knight shift spectrum of 
bulk sites for this new configuration. The distribution (thick line) is
broader than the corresponding homogeneous case (thin line). The reason is
the same as the one discussed in connection with Fig. 1 provided that the
relation $\Delta _{{\rm surf.}}/\Delta _{{\rm bulk}}$ is now inverted.
However, the NMR spectrum of surface
sites (Fig. 5) exhibits a considerable over-broadening, unlike the case presented 
in Fig. 3. This is due to the fact that, because of the topology of surface sites, completely
surrounding bulk sites, there is a strong mixing between surface states and
interior states, which also produces two fluctuation scales on the LDOS distribution
at the surface.

We now discuss the surface effects on the scaling law. Fig. 6 shows the relative
Knight shift as a function of the scaling parameter $\alpha=\eta/\Delta$. The
solid lines correspond to our results for particles of various sizes
with surface energies on all faces of the cube. 
These curves correspond to particles of $7\times 7\times 7$,
$10\times 10\times 10$ and $12\times 12\times 12$ sites.  
Experimental results by Kobayashi \cite{Kobayashi} \emph{et al} on aluminum
nanoparticles are also shown. The qualitative dependency of the relative knight
shift with $\alpha$ agrees very well with our result. In addition, a deviation of
the scaling law is clearly observed in both, simulations and experiment. 
Large particles lie further to the right than small particles. This can be 
interpreted as follow: due to the existence
of surface states, the mean spacing level at the Fermi energy is no longer $\Delta$,
since this last spacing is defined according to the density of states of the bulk. 
Therefore, $\alpha$ is no longer the scaling parameter. If we define $\Delta'$ as
the mean spacing level at the Fermi energy in presence of surface states and
$\alpha'$ the corresponding scaling parameter, then, according to Eq.~\ref{alpha}, 
$\alpha=\alpha'\frac{\Delta'}{\Delta}$. 
Since bulk states and surface states become uncorrelated and consequently the 
repulsion of levels around the Fermi energy decreases, $\Delta'<\Delta$. Therefore,
the factor $\frac{\Delta'}{\Delta}$ counts for the deviation of the scaling law. 
It is smaller for smaller particles than for larger particles which is consistent
with the experimental and predicted results in Fig. 6. 

\section{Conclusions}

According to our results, we expect that the NMR spectra of either surface
or bulk nuclei in small metallic nanoparticles, manifest the existence of
surface states as an over-broadened line which could also be interpreted as a
strong \emph{disorder}. In fact, the deviation of the scaling law shown in  
the experiment on aluminum nanoparticles \cite{Kobayashi} is an indication of
different distributions of energy spacings coming from bulk and surface states.
Additionally to the over-broadening of crystalline Ni
particles \cite{Brom} which is consistent with this view, there are other
ways in which these effects might be observed and tested: the 
NMR technique SEDOR (spin echo double resonance) consists of
simultaneous irradiation of resonances of nuclei of two different species
which are coupled either by direct dipolar interaction or through the
conduction electrons. With this method the contribution to the resonance
spectra of the nuclei from the surface can be inferred. A very interesting
application \cite{SEDOR} is the test of Platinum nanoparticles using Carbon
as a local probe. The last is chemisorbed on the Pt surface as CO. There,
the presence of a second peak in the Pt SEDOR\ line might be interpreted as
the coexistence of surface and bulk states.

Also based on the CO chemisorption at the surface of a metallic
nanoparticle, it has been observed \cite{Becerra,Shore,Zilm} that Carbon
NMR presents a shift toward a metallic nature. Not only the Korringa
equation for the relaxation time $T_{1}$ is verified ($1/T_{1}\propto
(N(\varepsilon _{{\rm F}},{\bf r}))^{2}/k_{B}T$) but the line shape presents
the inhomogeneity proper of quantum size effects. That is, due to the
mixing between the conduction band of the metal and the CO molecular
orbitals, the frequency shift of the C nuclei contains information of the
LDOS on the surface nuclei . Eventually an NMR study of chemisorbed molecules
also would give evidence of anomalous fluctuations in surface LDOS.

\section{Acknowledgments}
This work was performed in part at LANAIS de RMN (UNC-CONICET) with financial
support from SeCyT-UNC, CONICOR, CONICET and FUNDACION\ ANTORCHAS. The
authors thank discussions with V. N. Prigodin, H. B. Brom, K. Efetov, A.
Bonivardi, M.V. Ganduglia-Pirovano and J. Guevara as well as correspondence
with J. J. van der Klink and L. R. Becerra.

\end{document}